# Unusual rainbows as auroral candidates: another point of view

Víctor M. S. Carrasco[1], Ricardo M. Trigo[2], José M. Vaquero[1,3]

[1] Departamento de Física, Universidad de Extremadura, Badajoz, Spain (vmscarrasco@unex.es)
[2] Instituto Dom Luiz, IDL, Faculdade de Ciencias, Universidade de Lisboa, Lisbon, Portugal
[3] Instituto Universitario de Investigación del Agua, Cambio Climático y Sostenibilidad (IACYS), Universidad de Extremadura, Badajoz, Spain

**Abstract.** Several auroral events that occurred in the past have not been catalogued as such due to fact that they were described in the historical sources with different terminology. Hayakawa et al. (2016) have reviewed historical oriental chronicles and have proposed the terms "unusual rainbow" and "white rainbow" as candidates to auroras. In this work, we present three events that took place in the 18th century in two different settings (Iberian Peninsula and in Brazil) that were originally described with similar definition/wording used by the oriental chronicles, despite the inherent differences in terms associated to oriental and Latin languages. We show that these terms are indeed applicable to the three case studies from Europe and South America. Thus, the auroral catalogues available can be extended for occidental sources with this new terminology.

## 1. Introduction

In recent decades, a growing number of works have been devoted to characterize solar activity spanning several millennia. Solar activity and associated phenomena can be reconstructed through natural proxies (e.g. radionuclides in ice cores) but more often based on old documents (Usoskin, 2013; Vaquero & Vázquez, 2009). Solar variability can manifest in several ways, for example, through the appearance of sunspots on the solar disc or auroras in the Earth's atmosphere. Although there are records of naked-eye sunspot observations prior to the telescopic era, the invention of the telescope marked the beginning of the systematic measures of this phenomenon (Vaquero, 2007). The number of sunspots presented on the solar photosphere has an 11-year periodicity, approximately, and it is known as a solar activity cycle (Hoyt & Schatten, 1998; Clette et al., 2014). On the other hand, auroras (more common at high latitudes) are one of the most spectacular phenomena of nature, having always aroused a wide interest in all civilizations since many centuries ago (Brekke & Egeland, 1983; Feldstein et al., 2014; Akasofu, 2015). Unlike sunspots, observations of old auroras did not instigate obvious links with the contemporaneous solar activity and only recently such connection has been well established (Vázquez et al., 2014; 2016). In fact, auroras have a direct relationship with the solar activity cycle and therefore this kind of records are of a great interest due to the possibility of their use as a proxy of the solar activity in past times (Vázquez et al., 2016). Thus, long-term solar variability series, as sunspot or auroras, are interesting for various scientific fields considering the great influence of the Sun on several components of the Earth's Climate system (Peixoto & Oort, 1992) and our surrounding space (Usoskin, 2013; Clette et al., 2014).

Previous lists of medieval Chinese auroras have been compiled on the basis of several terms (Keimatsu, 1973; Xu, Pankenier & Jiang, 2000) as, for example, "red vapor" (Kanda, 1933; Yau, Stephenson & Willis, 1995). However, more recently it has been shown that such an approach can miss auroral events (Hayakawa et al., 2015). These latter authors employed a database of Chinese official chronicles to carry out a digital search using keywords as "vapor", "cloud" or "light" in order to find auroral events. In the chapter of "red vapors" of the Astronomical Treatise of Jiùtángshū (covering the Tang dynasty between 618 AD and 907 AD), they found the term "unusual rainbow" that, after careful analysis, was considered as an alternative description of an aurora (Hayakawa et al., 2016). Moreover, they found records including the term "white rainbow" which previously were not candidates to aurora because these are typically related to solar halos. However, after reading some of the "white rainbow" descriptions, these appear to be more consistent with an aurora definition. Note that the term "white rainbow" was also discussed before by Chapman et al. (2015) or Neuhäuser & Neuhäuser (2015) as auroral events. Taking all this terminology into account Hayakawa et al. (2016) proposed to look in further detail into the appearance of auroras descriptions using other key words besides the standard terms such as "vapor", "cloud" or "light". Thus, they have collected information about events that included the terms "unusual rainbow" and "white rainbow", using a mechanical retrieval system available for the digital database Scripta Sinica (http://hanchi.ihp.sinica.edu.tw/ihp/hanji.htm). After retrieving all the events that included these new key words the authors examined all of them very carefully to assess scientifically if those terms can be considered candidates to auroral event. Thus, according to these authors additional terms such as "unusual rainbow" and "white rainbow" appeared in this historical oriental sources as strong candidates to auroras. Here, we present three similar cases where these terms are also employed and that occurred in completely different parts of the globe. In particular, the events showed here were observed in the Iberian Peninsula and Brazil. Section 2 is devoted to introduce the cases with their original descriptions. In Section 3, we discuss the descriptions by the observers of these events and the main conclusions are presented in Section 4.

## 2. Observations

We have found three cases of nocturnal rainbow which, in our view, are strong candidates to auroras and, therefore, were not originally catalogued in this way. These events were observed in Iberian Peninsula and Brazil during the 18$^{th}$ century.

The first case analyzed was observed in the city of Porto (41º 9' N, 8º 36' W), northern Portugal, on 16 November 1729 (lunar phase: last quarter; geomagnetic latitude: 69º 44' N). The description of this event is included in a letter located in the Lisbon National Library that António Cerqueira Pinto sent to José Barbosa explaining the sighting of a possible comet [Cerqueira, A., 1729, Carta dirigida por António Cerqueira Pinto a D. José Barbosa sobre assuntos respectantes á sua orden e relatando o aparecimento possivel de um cometa. Porto (Portugal), 1729 November 19. Paper, 1 page, original handwritten, Codex 245, nº 133]:

ORIGINAL TEXT: "[…] *Na noite de 16 do corrente pellas 9 as dez horas apareceo nas vezinhacas desta Cid.$^e$ abanda entre norte e Occidente hum Fenomene dizem que do feitio de hua grande Raya de Corpo pouco claro entre nuvem espeça estando o Orbe sereno e claro, e que tinha abanda de fogo e que observara hum Religioso que então se vinha Recolhendo a Cid.$^e$ que estava atravesado de espadas de fogo e que finalizara en hua faixa o circulo de fogo a modo de arco Iris*".

ENGLISH TRANSLATION: "[…] During the night of 16th about 9h. to 10h. a band between north and West, it is said that a phenomenon appeared in the vicinity of this city that the shape of a big body line slightly clear among thick cloud being the sky quiet and clear, and it had a fire band which a religious who was arriving to this city observed as fire swords and finalizing in a band or circle of fire as a rainbow".

The second event took place in Rio de Janeiro (then capital of Brazil), 22º 54' S, 43º 12' W, on 20 June 1787 (lunar phase: first quarter; geomagnetic latitude: 18º 59' S). The description was extracted from *Memorias de Mathematica e Physica da Academia R. das Sciencias de Lisboa* (Sanches Dorta, 1812, p. 119):

ORIGINAL TEXT: "*No dia 20 de Junho [1787] esteve o Ceo muito limpo, e sereno conservandose o calor entre os 73.º e 74.º; e o Barometro en 28." 2'", 5. A' noite eu vî, e mostrei a dous sujeitos que estavão commigo, hum Arco Iris produzido pela refracção dos raios da Lua, que tinha as suas bazes ao Oest-sudueste, e Les-sueste, comprehendo todo o hemispherio. A sua côr era muito branca, sem que houvesse mistura de outra alguma côr. Começou a formar-se este arco ás 6.h 10' da noite, e acabou de formar-se ás 6.h 40' tempo de sua maior magnitude, cuja largura seria ao parecer dos nossos olhos d'huma braça: desappareceu este arco ás 7.h 18' da noite. No intervallo da sua duração foi-se movendo para o Sul, e lá acabou.*

*A Lua estava clarissima, e contava-se o seu sexto dia: tinha passado pelo nosso Meridiano ás 4.h 39' da tarde: achava-se com Declinação Boreal de 3.º 32' com pouca differença, e no signo de Virgo.*

*Toda a atmosfera estava alguma cosa avermelhada, e quando o arco desappareceu, encheu-se d'huma nuvem muito branca para a parte do Sul. O Vento assoprava muito brando da parte do NE: e relampejava da mesma base do arco Les-sueste, etc*".

ENGLISH TRANSLATION: "On June 20 [1787] the sky was very clear and quiet keeping the heat (temperature) between 73º and 74º; and the barometer at 28." 2'", 5. During the night I saw, and I showed to two individuals who were with me, a rainbow produced by refraction of rays of the Moon, which had its bases on the west-southwest, and east-southeast occupying completely the hemisphere. Its color was very white, with no mixture of any other color. It started to form this arch at 6.h 10' of the night, and ended up at 6.h 40', time of its greater magnitude, whose width would be by our eyes one fathom [1.8m approximately]: it disappeared this arc at 7h 18' of the night. In the interval of its duration was moving towards the south, and there it ended.

The moon was very clear, and it was into its sixth day: it had passed by our meridian at 4h 39' in the afternoon: it found with boreal declination 3º 32' with little difference and in the sign of Virgo.

The whole atmosphere was slightly reddish, and when the arch disappeared, it filled of a very white cloud for the Southern sector. The wind blew softly from the NE sector: and it flashed [lightning] of the same base of the east-southeast arch, etc".

Finally, the last of the cases presented here occurred in Jaén (south of Spain) on 17 June 1788 (lunar phase: full moon; geomagnetic latitude: 64º 2' N), 37º 46' N, 3º 47' W, and was published in a former Spanish newspaper *Memorial Literario, Instructivo y Curioso de la Corte de Madrid* (1788, Tomo XIV, p. 261):

ORIGINAL TEXT: "Uno de nuestros Corresponsales Literatos residente en la Ciudad de Jaen nos ha comunicado que en la noche del dia 17 de este mes en que habia habido una pequeña tormenta vió formado un arco iris muy perfecto, con solo la diferencia de tener un color como ceniciento; ignora á la hora que tuvo principio, pero desde que lo vió duró como un medio quarto de hora, cuya situación seguia unos cinquenta grados ácia el Norte".

ENGLISH TRANSLATION: "One of our writer correspondents resident in the city of Jaen told us that the night of the 17th of this month in which there had been a small storm saw a perfect rainbow being formed, with only the difference of having a color like ashy; he ignores the initial hour, but since he saw it lasted about a half quarter hour, whose situation was about fifty degrees north".

## 3. Discussion

First, we must stress that the three events described here occurred during dark night and the hours established for each one are given by the local time (counted from midday). Furthermore, we have consulted the auroral catalogues by Fritz (1873) and Angot (1896) in order to find auroral events in other places for the same dates found in the events described in this work. In these catalogues, we can find: i) several auroral sightings for the same date of the first event registered, for example, in Leipzig or Paris; ii) one auroral record in Cambridge (one day before) and another one in Rome (one day after) of the second event and iii) there are no records near the date of the third event. On the other hand, we have checked the recently revised collection of sunspot group numbers (Vaquero et al., 2016) to know if sunspot records exist for the dates when these events occurred. Unfortunately, there are no sunspot records for these days. However, we note that these three events occurred during maxima of the Schwabe cycle. Furthermore, we have added the geomagnetic latitudes of the places for the corresponding years of the events using the gufm1 geomagnetic model (Jackson et al., 2000) and the web-application of the National Geophysical Data Center (http://ngdc.noaa.gov/geomag-web/#igrfwmm).

We acknowledge that the description of the records studied here could be confused with other optical phenomena produced in the atmosphere such as the noctilucent clouds, fog bow and lunar rainbow or halos. However, taking into account all the information gathered we are relatively confident that these three observations correspond, in fact, to auroras for the reasons that we elaborate further below.

On the one hand, noctilucent clouds are visible in the twilight and, generally, in places located over high latitudes. In our cases, all the three events developed during the night and the three cities where these were observed are located at mid-latitudes. We can almost discard the fog bow for the three events because the observers do not indicate the presence of fog although we note that Minnaert (1993) indicated fog bow has also been seen when the fog was so thin that the observer who saw it declared that there was no fog. On the other hand, lunar halos appear, as their names suggest, around the moon. In particular, the moon is usually located in the southern part of the sky for the places of the northern hemisphere (and in the northern part of the sky for southern hemisphere places). However, our events are observed on the opposite direction, i.e. in the northern part of the sky for Portugal and Spain and in the southern part for Brazil. Moreover, the description of the observer from Rio de Janeiro points out that the phenomenon encompasses the entire sky. Moreover, lunar rainbows can be rejected in the cases of Portugal and Brazil due to the fact that the sky was clear. For the event occurred in Spain, a lunar rainbow can be possible because there was a small thunderstorm although we think that the most likely phenomenon is an aurora.

On the other hand, the color of the rainbow observed in Spain and Brazil are defined as grayish and white, respectively. The colors of the auroras hold some information about the intensity of the auroras (Hayakawa et al., 2015). The most frequent colors of the auroras are green, red, violet, etc., but rarely white. Nevertheless, when an aurora is not very bright the human eye tends to recognize it as a white color (Hayakawa et al., 2016) and that was probably what happened also in our three events. Moreover, we highlight that the rainbow observed in the chronicles of Porto is defined as a circle of fire discarding, thus, the possibility that this observation is a comet and the observer in Rio de Janeiro pointed that atmosphere was a little reddish.

We have checked for the three cases which auroral criteria are fulfilled according to five criteria defined by Neuhäuser & Neuhäuser (2015). Thus, for the first event (16 November 1729, Portugal), four auroral criteria are fulfilled (colour, auroral motion, northern direction and night-time observation) and, therefore, this event is classified as a "very probable" aurora according to those criteria. The case of the second event (20 June 1787, Brazil) is the same because four criteria are fulfilled (colour, auroral motion, southern direction and night-time observation) and, so, this event is also classified as a "very probable" aurora. Last, for the third event (17 June 1788, Spain), unlike the other cases, two criteria are fulfilled (northern direction and night-time observation) and it can be considered as a "very possible" aurora. Note that although, according to these criteria, we have considered that an auroral event is "very probable" or "very possible", we have checked in this section if these events are consistent with other phenomena.

Finally, we acknowledge that the scientific knowledge of the three observers involved varies considerably and is difficult to evaluate. Nevertheless, it is important to stress the outstanding scientific knowledge and quality of observations provided by Sanches Dorta in Brazil, as noted in many previous studies (Vaquero & Trigo, 2005; 2006; Trigo et al., 2010; Farrona et al., 2012).

## 4. Conclusions

The meticulous astronomical observations and descriptions undertaken by medieval Chinese scholars have allowed the establishment of long-term lists of auroras often based on the term "red vapor" (e.g. Yau, Stephenson and Willis, 1995). Recently, Hayakawa et al. (2015) have improved the search for terms related to auroral events employing full automated digital search based on the keywords "vapor", "cloud" or "light". However, Hayakawa et al. (2016) have showed that some auroral events could have been registered with other words. Thus, after studying several Chinese official chronicles, they have proposed the terms "unusual rainbow" or "white rainbow" as candidates to auroral events. We present other three cases that in our view represent strong candidates to auroras observed in other places around the world, including two in Iberia (Portugal and Spain) and a third one in Brazil. After analyzing these events, we are confident that all three case studies correspond to real auroras and that the observers used the term "rainbow" only because they were not able to identify the auroral phenomenon. Therefore, the seminal idea raised by Hayakawa et al. (2016) can be extended for other historical sources around the world. If that is doable then the number of records compiled in the auroral catalogues could be increased significantly, particularly in Europe and America, where this type of assessment has not been performed so extensively. This preliminary work presented here aims to contribute in that direction, although only a few examples are shown based on manual identification, i.e. without employing an automatic search of key words in electronic texts.

**Acknowledgements**
Support from the Junta de Extremadura (Research Group Grants GR15137) and from the Ministerio de Economía y Competitividad of the Spanish Government (AYA2014-57556-P) is gratefully acknowledged.